\documentclass[referee]{raa}           
\usepackage{graphicx,times}
\usepackage{natbib}
\usepackage{amssymb,amsmath}
\usepackage{amsmath}
\usepackage{url}
\usepackage{graphicx,graphics,amsmath}
\usepackage{subfig}
\usepackage[T1]{fontenc}
\usepackage[utf8]{inputenc}
\usepackage{lipsum}

\newcommand{\Om}{\Omega_{\rm M}}

\newcommand{\Ox}{\Omega_{\rm X}}

\bibpunct{(}{)}{;}{a}{}{,}

\usepackage[a4paper=true,dvipdfm=true,pagebackref=true]{hyperref}
\hypersetup{pdftitle = The title of my PDF, pdfauthor = My name, pdfsubject= The subject, pdfkeywords = keyword1 keyword2 keyword3} 
\hypersetup{colorlinks = true, linkcolor = green, anchorcolor = red, citecolor = blue, filecolor = red, pagecolor = red, urlcolor = red}

\begin{document}

   \title{On The Non-Gaussian Errors in High-z Supernovae Type Ia Data
}

 \volnopage{ {\bf 2016} Vol.\ {\bf X} No. {\bf XX}, 000--000}
   \setcounter{page}{1}

   \author{Meghendra Singh\inst{1}, Ashwini Pandey\inst{2}, Amit Sharma\inst{2}, Shashikant Gupta
      \inst{2}, Satendra Sharma\inst{3}
   }

   \institute{Dr.A.P.J.Abdul Kalam Technical University, Uttar Pradesh, Lucknow 226021, India; {\it meghendrasingh\_db@yahoo.co.in}\\
	\and Amity University Haryana, Gurgaon, Haryana 122413, India.\\
	\and Yobe State University, Damaturu, Yobe State, Nigeria.\\ 
\vs \no
   {\small Received----- ; accepted -----}
}

\abstract{The nature of  random errors in any data set is Gaussian is a well established fact
according to the Central Limit Theorem. Supernovae type Ia data have played a crucial role in 
major discoveries in cosmology. Unlike in laboratory experiments, astronomical measurements can not
be performed in controlled situations. Thus, errors in astronomical data can be more severe in 
terms of systematics and non-Gaussianity compared to those of laboratory experiments. 
In this paper, we use the Kolmogorov-Smirnov statistic to test non-Gaussianity in high-z supernovae data. 
We apply this statistic to four data sets, i.e., Gold data(2004), Gold data(2007), Union2 catalogue and the Union2.1 data 
set for our analysis. Our results shows that 
in all four data sets the errors are consistent with the Gaussian distribution. 
}

   \authorrunning{M.Singh et al. }            
   \titlerunning{On The Non-Gaussian Errors in High-z Supernovae Type Ia Data}  
   \maketitle

%
\section{Introduction}           
\label{sect:intro}

The light curves of Type Ia supernova (SNIa) have been used as cosmological distance indicators 
(\citealt{Riess+1998,Perlmutter+1999}) to mark out the expansion history and to detect cosmic acceleration 
as well. The overall picture of the Universe is consistent with a model known as the $\Lambda$CDM, consisting 
of around one quarter of baryonic and dark matter and three quarters of dark energy. The dark energy 
can be treated as a cosmic-fluid with equation of state $P=w\rho$; where the pressure ($P$) is allowed 
to be negative. The SNIa data can be used to constrain the equation of state parameter ($w$) which is 
the key to study dark energy (\citealt{Freedman+2009,Hicken+2009,Rest+etal+2014,Scolnic+etal+2014})	. 

However, many alternative explanations exist for dark energy and its exact nature is also unknown. For 
instance, a classical fixed cosmological constant, $\Lambda$, yields $w = -1$, whereas other models 
(e.g. quintessence) yield values of $w > -1$ (\citealt{Huntere+etal+2001}). To overcome this difficulty, precise 
enough data is required to detect fluctuations in the dark energy. The data should also 
cover wide range of redshifts to constrain the detailed behavior of dark energy with time.
Presently, the data fulfilling the above criteria is obtained by the observations of the SNIa. 
Determination of supernovae distances having high precision and tiny systematic errors is crucial 
for above purpose; and we would like to be certain that their statistics is well understood. 
Further, if Central Limit Theorem holds (\citealt{Kendall+etal+1977}), the statistical 
uncertainties in SNIa data should follow Normal distribution. The systematics, if present, have to be 
identified and removed separately. Treatment of the errors becomes more important in astronomy 
since it is hard to repeat or perform the experiments in controlled way unlike the 
laboratory experiments. In the present paper, we use the Kolmogorov-Smirnov test (hereafter KS test) in an elegant way 
to detect the non-Gaussian uncertainties in SNIa data. 


This paper aims to address the above mentioned problems. The rest of the paper is formed as follows: 
In \S~2, we illustrate the different data sets used for our analysis , while \S~3 contains detailed description of methodology used. 
In \S~4, we continue and put forward our results for various data sets and lastly \S~5 is reserved for conclusions.

\section{Data}
\label{sec:data}
The Gold data GD04 (\citealt{Riess+etal+2004}) containing 157 SNe, GD07 (\citealt{Riess+etal+2007}) containing 182 SNe along with  
the more recent and larger data sets Union2 (\citealt{Amanullah+etal+2010}) and Union2.1 (\citealt{Suzuki+etal+2012})  
containing 557 and 580 SNe respectively are used to carry out our investigation.  The redshift $z$ and the distance 
modulus $\mu$ are the measured quantities in the data. If $m$ is the apparent magnitude and $M$ is the absolute 
magnitude, then distance modulus is defined as:
\begin{equation}
\mu(z) = m(z) - M\,, 
\label{eq:mu1}
\end{equation}
The apparent magnitude $m(z)$ and hence distance modulus $\mu(z)$ depends on the intrinsic luminosity of a supernova, its redshift $z$ and the cosmological parameters. The distance modulus $\mu(z)$ and the luminosity distance $d_L$ are related as:
\begin{equation}
\mu(z) = 5 \log \left({d_L(z)} \right) + 25\,,
\label{eq:mu}
\end{equation}
where the luminosity distance is measured in $Mpc$ and follows:
\begin{equation}
d_L(z) = \frac{c (1+z)}{H_0}\int_{0}^{z} \frac{dx}{h(x)}\,,
\label{eq:d_L}
\end{equation}
where $h(z; \Om,\Ox)=H(z; \Om,\Ox)/H_0$, and hence it is independent of $H_0$ but
depends purely on densities of dark matter ($\Om$) and dark energy ($\Ox$). 
The variation of $\Ox$ with redshift is already encoded in the cosmological models; for instance $\Ox$ is a constant in the $\Lambda$CDM model. The nature of relation of $\mu$ with $M$ is linear, however, that with luminosity distance is logarithmic. This implies the logarithmic dependence of $\mu$ on Hubble parameter $H_0$ as well.

\section{Methodology}
\label{sec:method}
We now give an introduction to the method of our analysis. Originally it was presented in 
\citealt{Singh+etal+2015}(hereafter GS15) to find non-Gaussianity in the HST Key Project Data.

If the correct theoretical value of the distance modulus of $i^{th}$ supernova at redshift $z$
is $\mu_i^{th}(z)$, then the observed value $\mu_i^{obs}$ will be 
\begin{equation}
\mu_i^{obs} = \mu_i^{th}(z) \pm \sigma_i ;
\end{equation}
where $\sigma_i$ is the error in the measurement of distance modulus. We expect these errors to be 
completely random, however, there could be some undesired contribution from systematic effects. For the time being
we assume that the systematic part in the errors is zero. We show in next paragraph that the
presence of systematic errors will not affect our analysis. Further, Central Limit Theorem suggests that
the random part of the errors should be Gaussian in nature with mean value zero.
Now we define a quantity $\chi_i$ such that:
\begin{equation}
\chi_i = \frac{\mu_i^{obs} - \mu_i^{th} (z)}{\sigma_i} ,
\label{eq:chi}
\end{equation}
Clearly $\chi_i$ should follow the standard normal distribution ~N(0,1), i.e., Gaussian distribution with
zero mean and unit variance. The effect of random errors is to scatter the data around the true value and that 
of systematics is to shift the average away from the true value. If the systematics are present they will just 
shift the average, hence one should subtract the best-fit value rather than true theoretical value in Eq.~\ref{eq:chi}.
Thus Eq.~\ref{eq:chi} takes the following form for a given SN:
\begin{equation}
\chi_i = \frac{\mu^{obs}_i - \mu_i^{bestfit}(z)}{\sigma_i} ,
\label{eq:chinew}
\end{equation}
where  $\mu_i^{bestfit} (z)$ is calculated using the best-fit values of cosmological parameters. 
Statistical independence among supernovae in our analysis is an obvious assumption. $\chi_i$ defined in Eq.~\ref{eq:chinew} 
should follow a standard normal distribution, i.e., Gaussian with zero mean and unit standard deviation.

\begin{figure}[tp]
\subfloat{
\includegraphics[width=7.0cm,height=7.0cm]{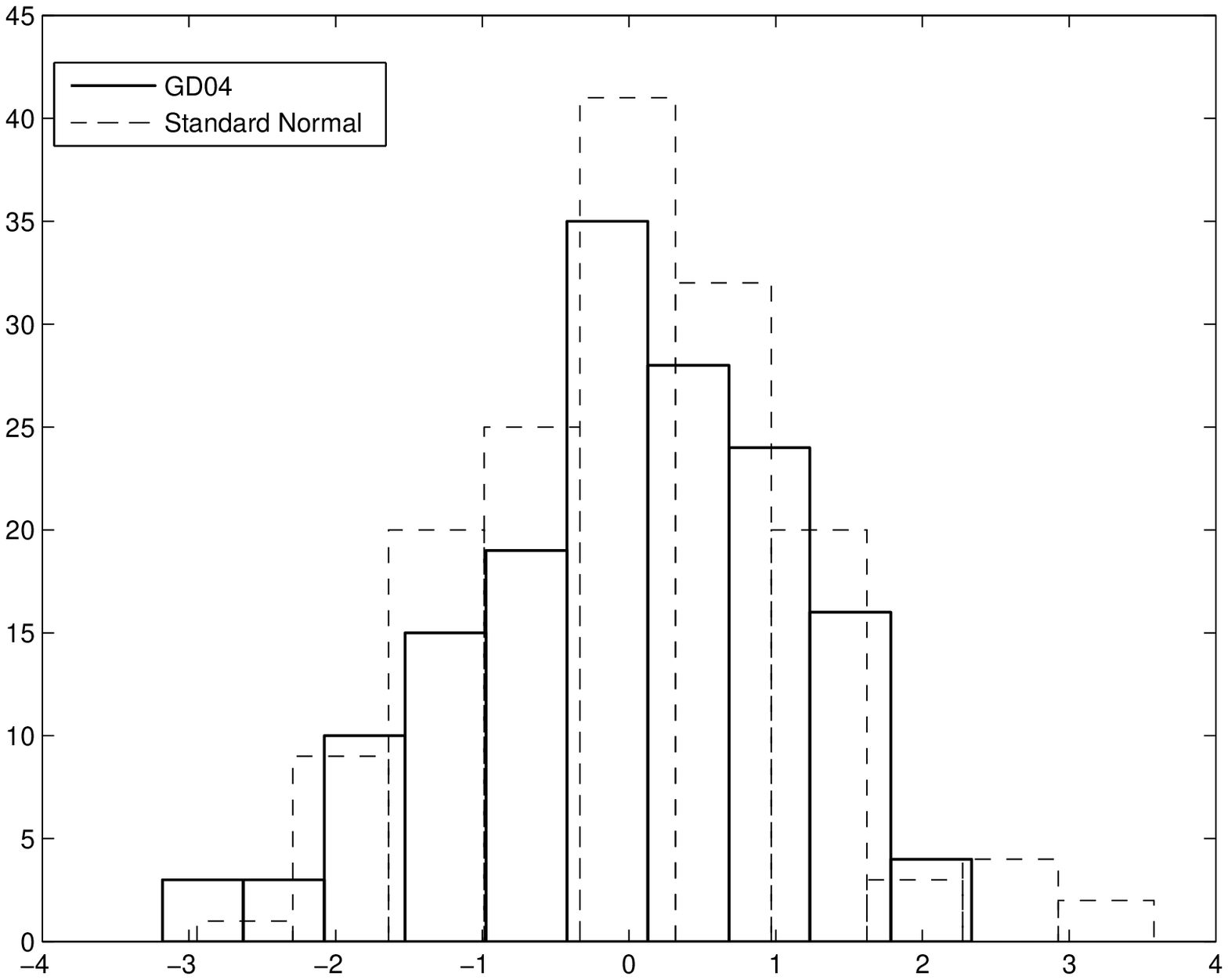}}
\subfloat{
\includegraphics[width=7.0cm,height=7.0cm]{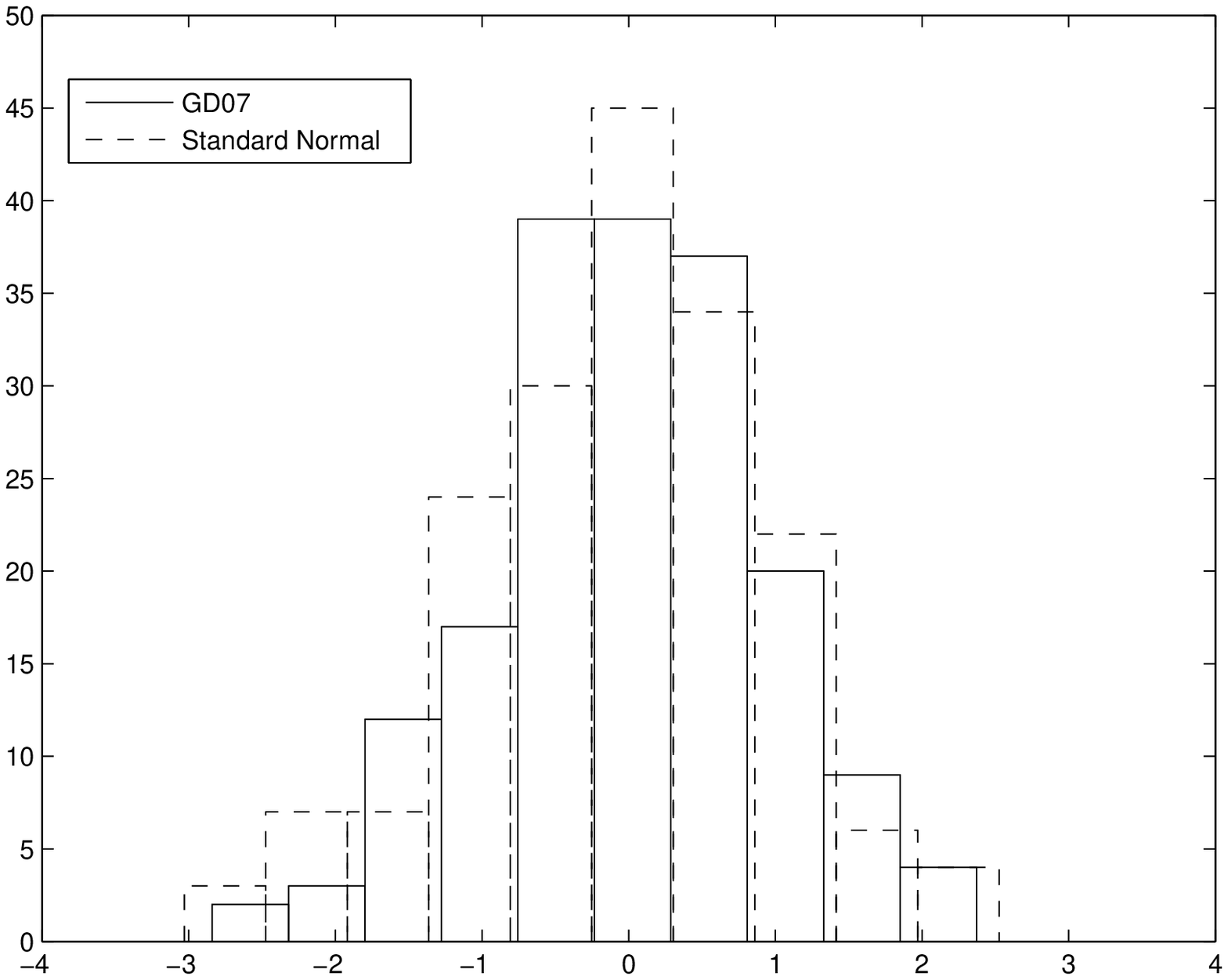}}\\
\subfloat{
\includegraphics[width=7.0cm,height=7.0cm]{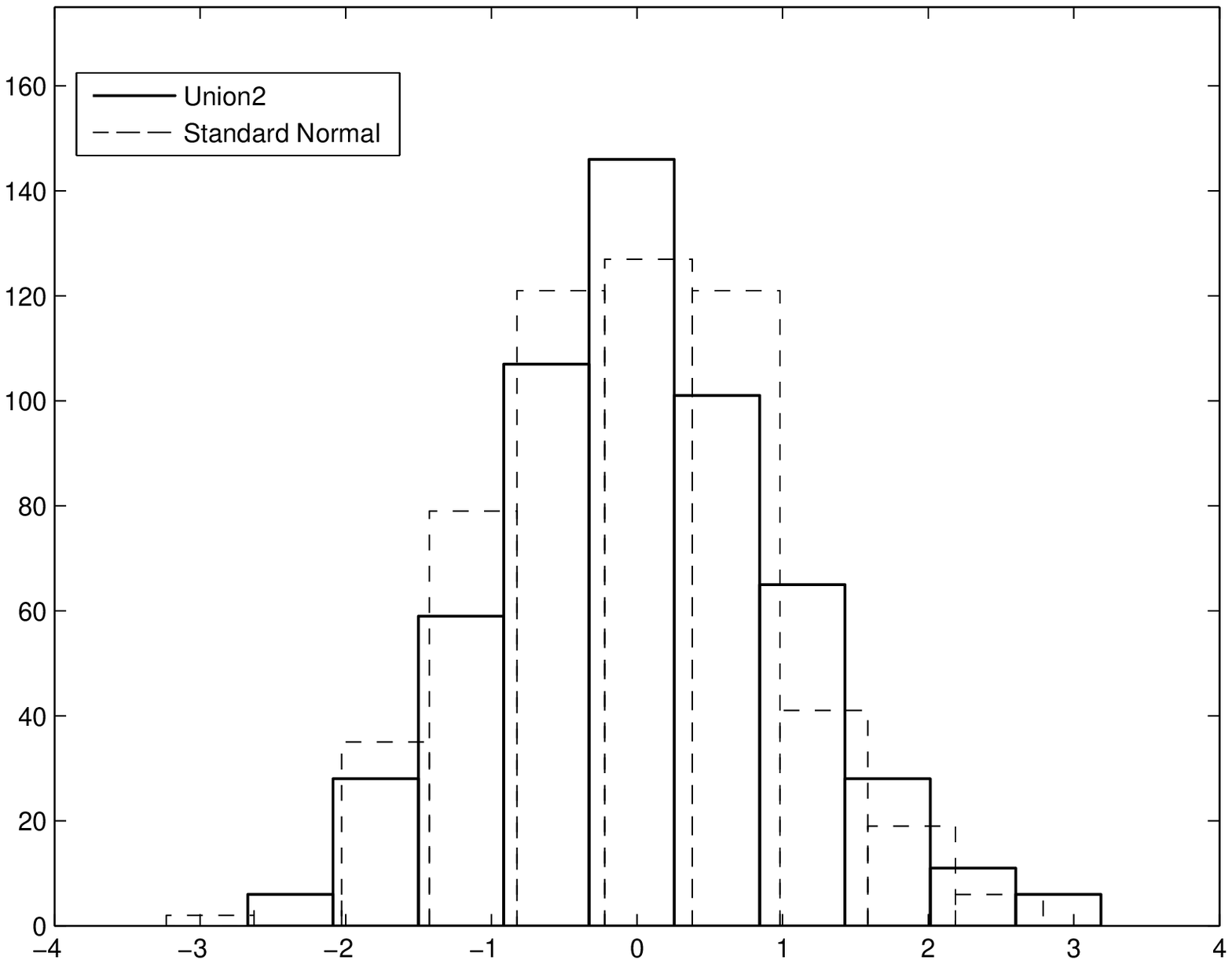}}
\subfloat{
\includegraphics[width=7.0cm,height=7.0cm]{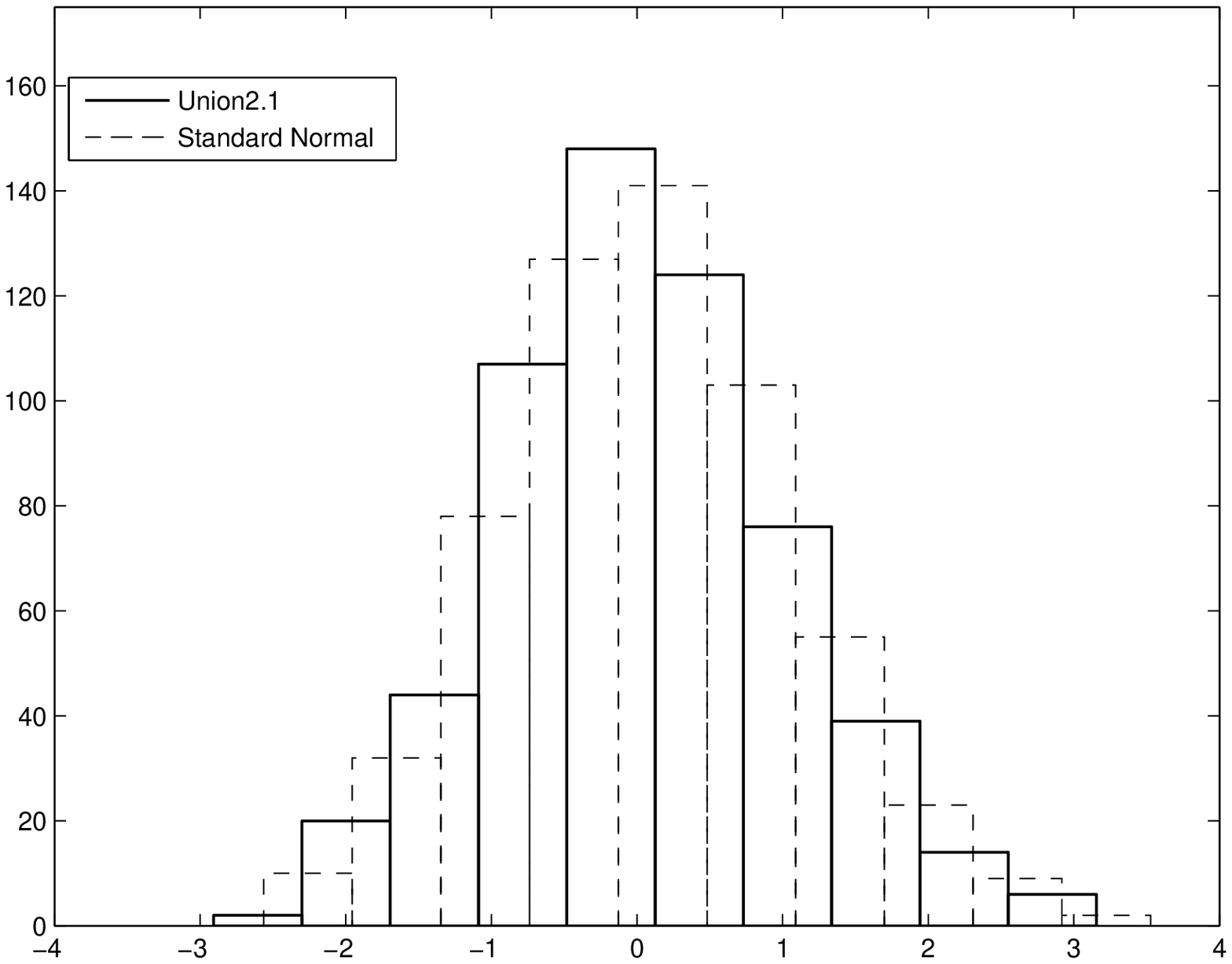}}
\caption{\label{fig:hist} Histogram of $\chi_i$ for various data set is compared with that of standard
normal distribution.}
\end{figure}

We use the flat $\Lambda$CDM cosmology in our analysis, since it fits the SNe data well. 
However, other cosmological models could also be investigated using a similar approach.
In order to get best-fit values of Cosmological parameters we minimize $\chi^2$, which is defined as:
\begin{equation}	
\chi^2 = \Sigma_{i=1}^N {\big[ \frac{\mu^i - \mu^{\Lambda CDM}}{\sigma_i} \big]}^2 \, ,
\label{eq:bestfit}
\end{equation} 
Once again we emphasize that, Eq.~\ref{eq:bestfit} is used to find the best-fit values of cosmological
parameters and it is then used in Eq.~\ref{eq:chinew} to calculate $\chi_i$.

As argued earlier, $\chi_i$ should follow standard normal distribution. To check this, we use KS test 
to determine whether or not a given sample follows the Gaussian distribution 
(\citealt{Press+etal+2007}). For this we set our null hypothesis as: "The errors in the SNe data are drawn from a Gaussian 
distribution". Thus $\chi_i$'s in Eq.~\ref{eq:chinew} would follow standard normal distribution. We apply KS test to 
calculate the test statistic and the p-value which is the probability of attaining the observed sample results when 
the null hypothesis is true. 

\begin{figure}[]
\subfloat{
\includegraphics[width=7.0cm,height=7.0cm]{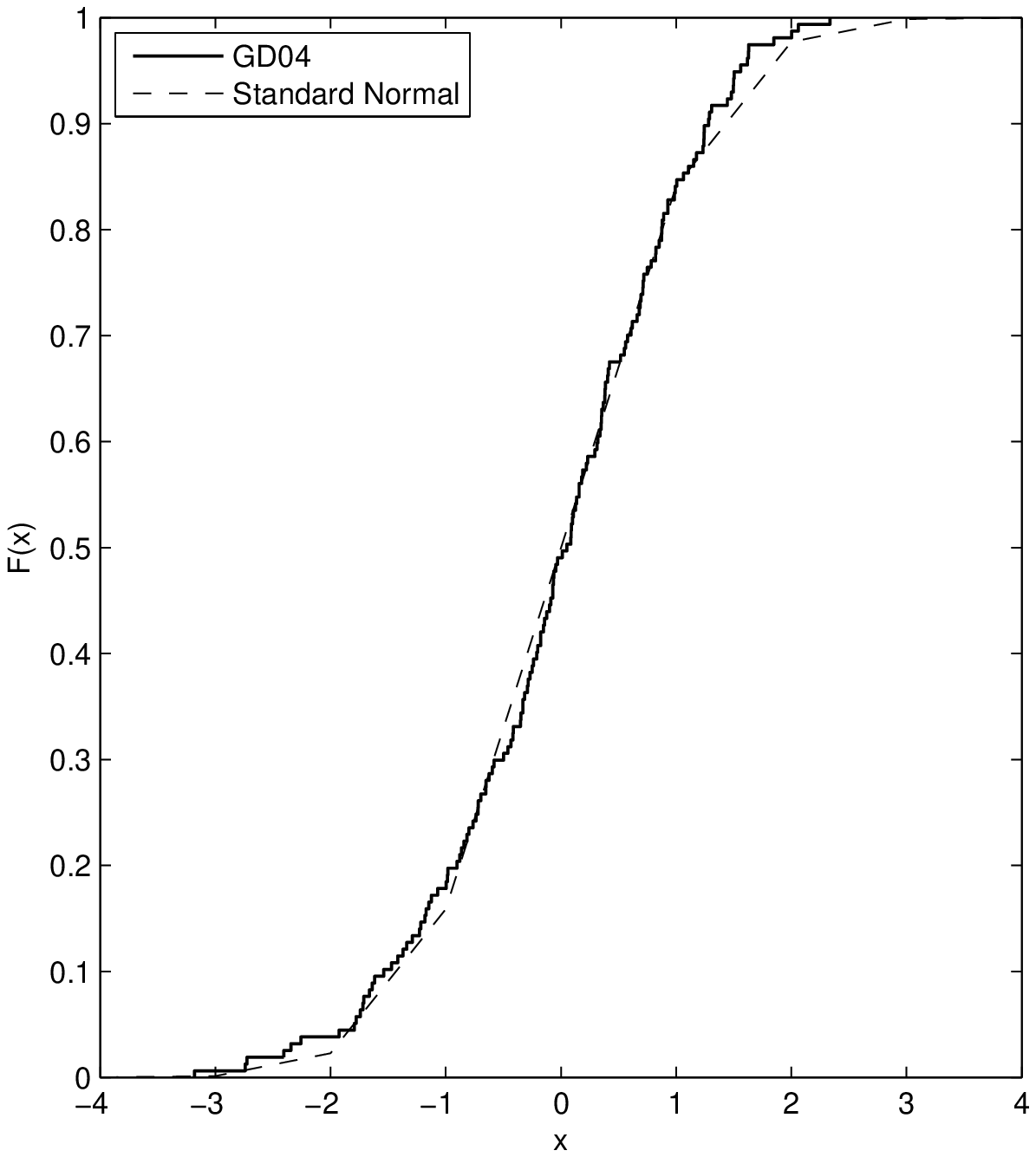}}
\subfloat{
\includegraphics[width=7.0cm,height=7.0cm]{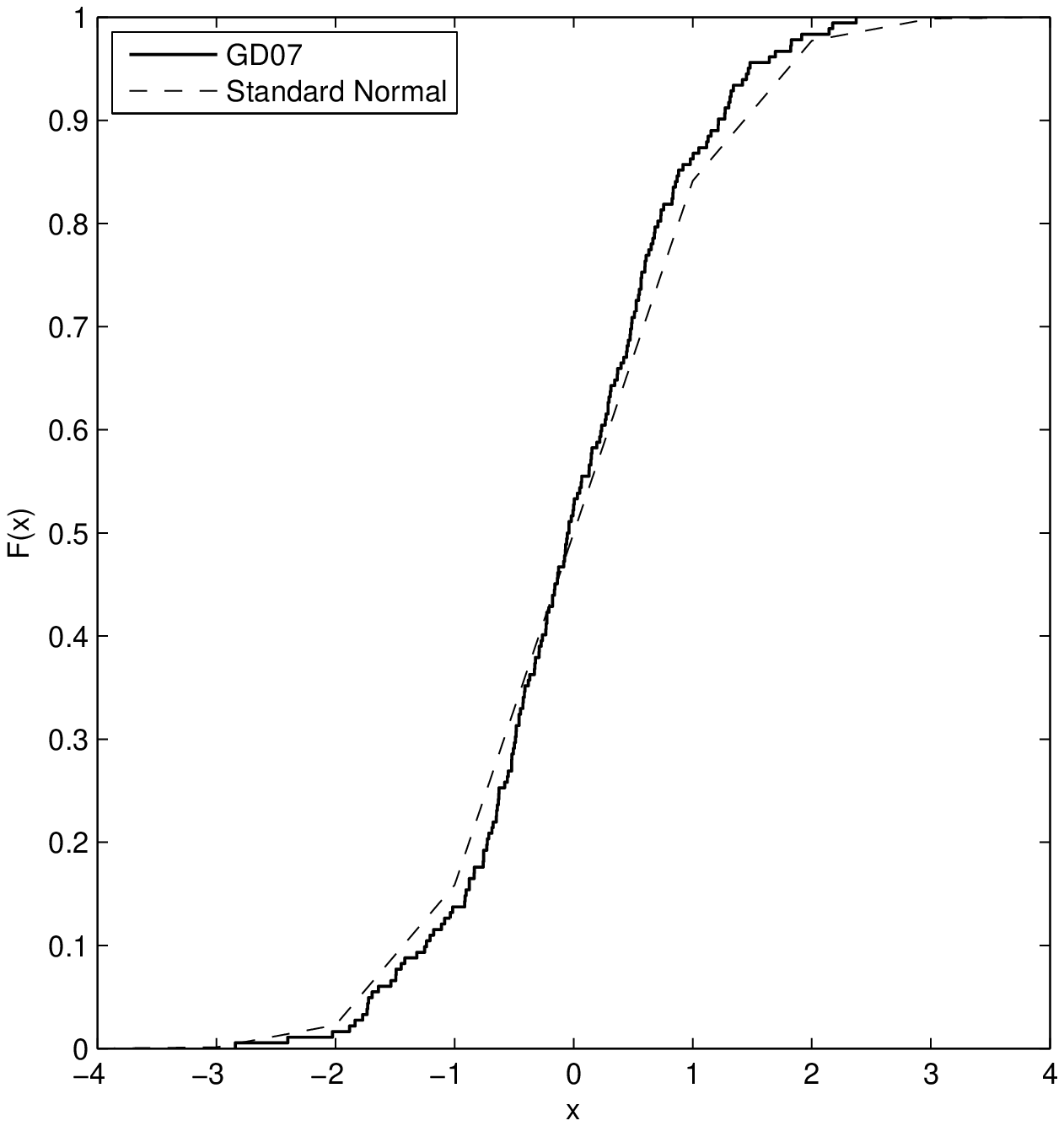}}\\
\subfloat{
\includegraphics[width=7.0cm,height=7.0cm]{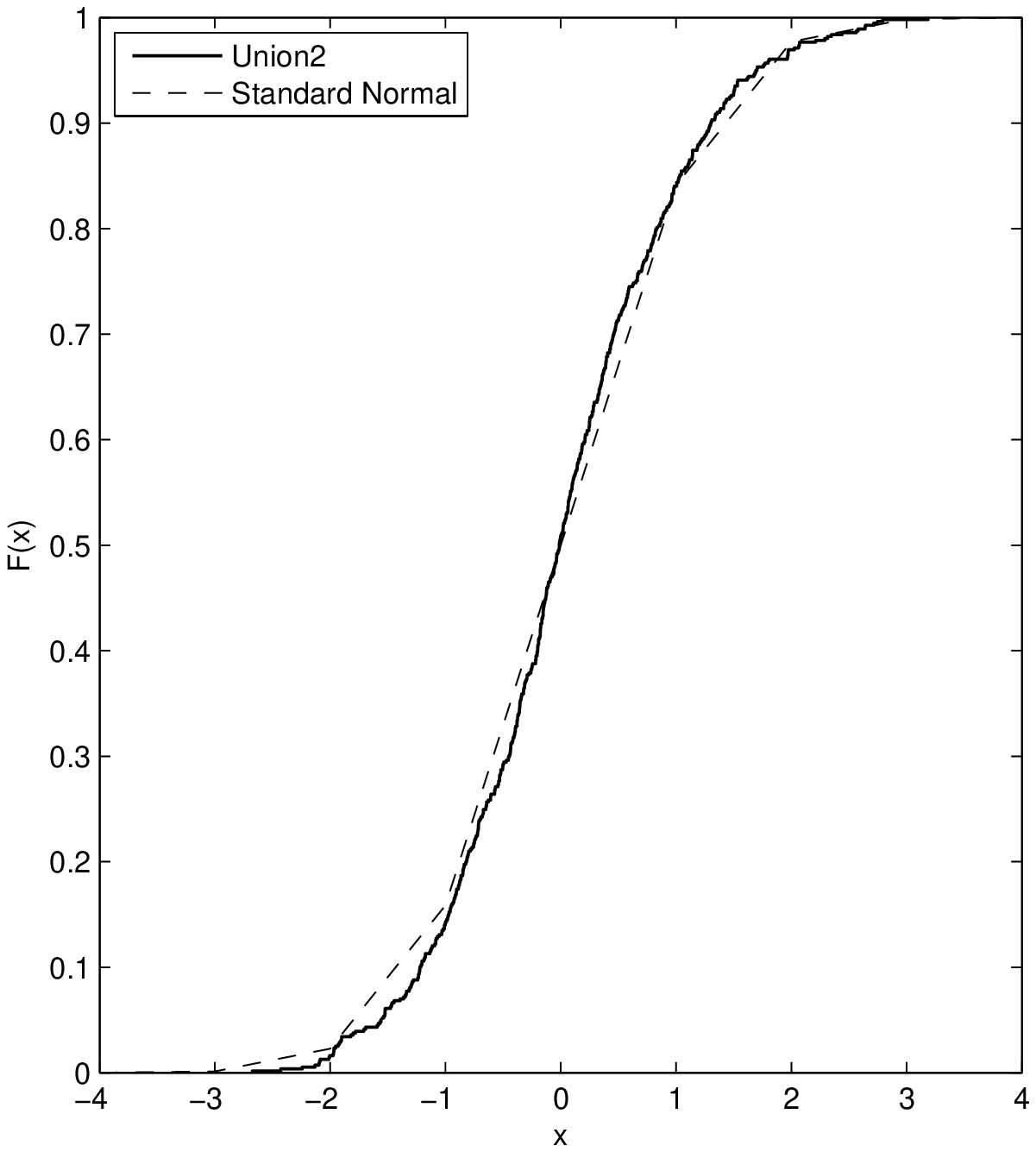}}
\subfloat{
\includegraphics[width=7.0cm,height=7.0cm]{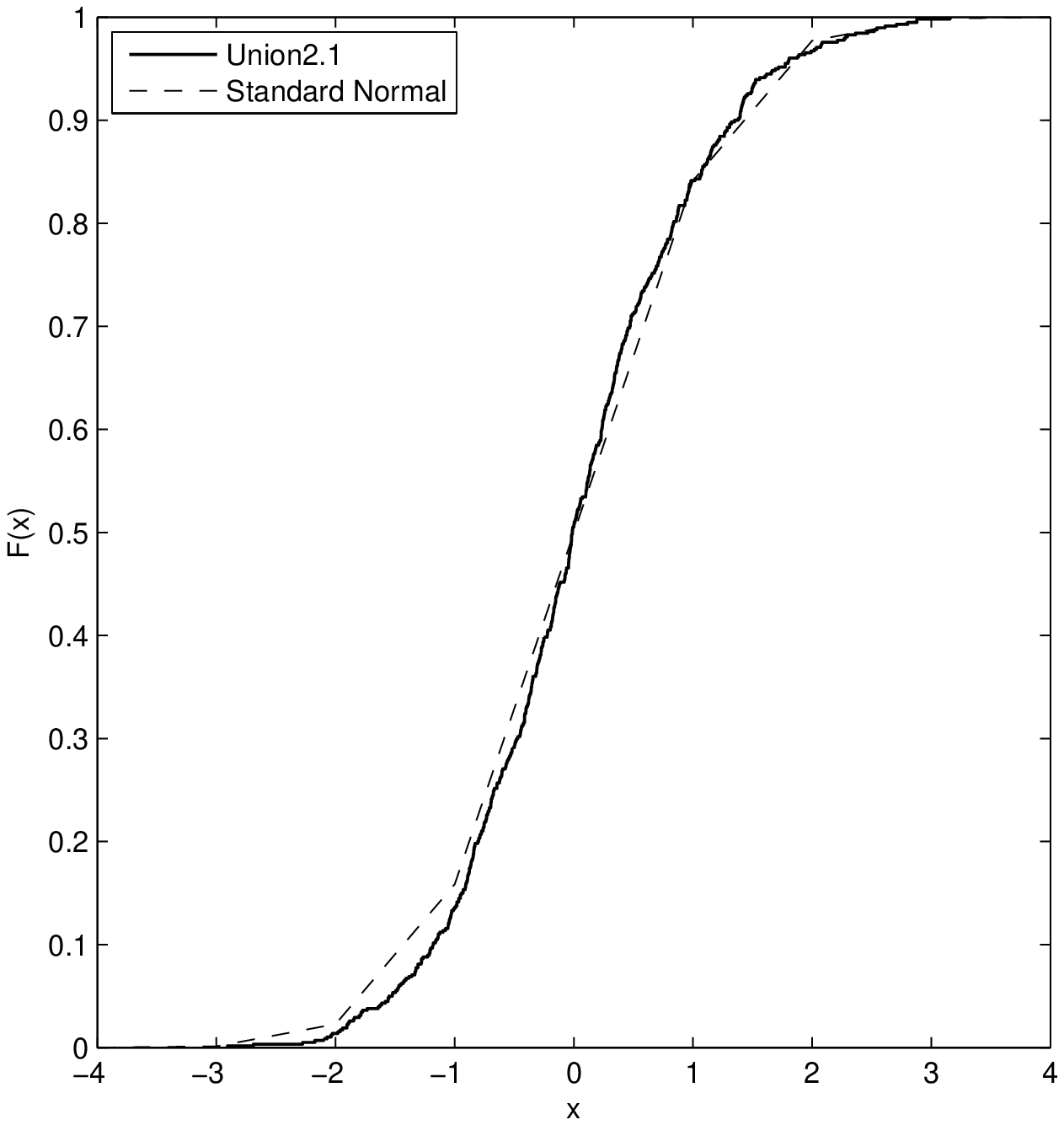}}
\caption{\label{fig:CDF} Comparison of cumulative distribution of $\chi_i$ for different data sets with their Gaussian CDF. 
Smooth curve represents the Gaussian CDF}
\end{figure}

For this, we use Matlab function \emph{kstest[h,p,k,cv]}; where: $p$ represents the probability of 
the data errors being drawn from Gaussian distribution, $k$ is the maximum distance between the 
two distributions (CDF), and $cv$ is the critical value which is decided by the 
significance level ($\alpha$). Different values of $\alpha$, indicate different tolerance 
levels for false rejection of the null hypothesis. For instance, $\alpha=0.01$ means that we 
allow 1\% of the times to reject the null hypothesis when it is true. $cv$ is the critical
value  of the probability to obtain/generate the data set in question given the null hypothesis; and 
is to be compared with $p$. 
A value $h=1$ is returned by the test if $p < cv$ and the null hypothesis is rejected.
While for $p>cv$, $h$ remains $0$ and the null hypothesis is not rejected. 

\begin{table}
\begin{center}
\caption{
  The best-fit values for various data sets.}  
  \label{tbl-bestfit} \bigskip
\begin{tabular}{ccrrr}
\hline
Data Set & \# SNe & $\Om$ & $ H_0 $ &  $\chi^2/dof$ \\
\hline
GD04 & 157 & 0.30 & 64.5 &  1.143 \\
GD07 & 182 & 0.33 & 63.0 &  0.883 \\
Union2 & 557 & 0.27 & 70.0 &  0.975 \\
Union2.1 &580 & 0.28 & 70.0 &  0.973 \\
\hline\hline                                                                                        
\end{tabular}
\end{center}
\end{table}

\section{Results}
We apply the statistic dissucssed in section~\ref{sec:method} on various SNe data sets and present the results here. Similar analysis was presented in \citealt{Gupta+etal+2010} (hereafter GS10) and in \citealt{Gupta+etal+2014} (hereafter GS14) using a different method ($\Delta\chi^2$) based on extreme value theory.

As a first check, we calculate the best-fit values of cosmological parameters for all four data sets by minimizing $\chi^2$ which are 
presented in Table~\ref{tbl-bestfit}. It is clear that both gold data sets favor higher matter density ($\Om$)
and consequently smaller expansion rate ($H_0$) compared to Union2 and Union2.1. One important fact is that the $\chi^2$ per degree of freedom secures smallest value for GD07 while largest for GD04, indicating the overestimation and underestimation of errors in GD07 and GD04 respectively.

We calculate $\chi_i$'s as defined in Eq~\ref{eq:chinew} for each data set using the best-fit values 
presented in Table~\ref{tbl-bestfit}. Further, we generate four sets of random numbers following 
Gaussian distribution with zero mean and unit standard deviation. Fig.~\ref{fig:hist} represents 
the comparison of histograms of Gaussian random numbers with that of $\chi_i$'s of each data set.

Secondly, the result of the KS test which is arrived at by comparing the calculated cumulative distributions for $\chi_i$'s with that of Gaussian distribution are presented in Table~\ref{tbl:ksresults}. The second, third and fourth column in Table~\ref{tbl:ksresults} denote values of $p$, $k$ and $cv$ respectively. Since $p>cv$ in all cases giving $h=0$, which means that we can not reject the null hypothesis that the errors are drawn from a Gaussian distribution; is shown explicitly by Fig.~\ref{fig:CDF}. 

\begin{table}
\begin{center}
\caption{
 Results of KS test for various data sets.}  
\label{tbl:ksresults} \bigskip
\begin{tabular}{crrr}
\hline
Data Set & $p$ value & $ k $ & $Cv$ \\
\hline
GD04 & 0.9280 & 0.0425 & 0.1073\\
GD07 & 0.7872 & 0.0475 & 0.0997\\
Union2 & 0.7328 & 0.0288 & 0.0572 \\
Union2.1 & 0.6764 & 0.0296 &  0.0561\\
\hline\hline                                                                                        
\end{tabular}
\end{center}
\end{table}

\section{Conclusions}
We have used the method presented in GS15 to detect non-Gaussianity in the error bars in Supernovae
data. Our main conclusions for this part of our work are following: 
(a) The errors are probably underestimated in GD04 and overestimated in GD07. In this sense, both the 
sets stand on extreme positions. 
(b) For a flat $\Lambda$CDM cosmology GD07 favors slightly higher matter density and this can 
be verified by the fact that in GD07 the distances are smaller compared to that in GD04 set for common supernovae. 
(c) Comparison with GS10 and GS14: GD04 was shown to have non-Gaussian component of errors while
in KS test, it shows the highest probability of being consistent with Gaussian distribution. (d) The hypothesis that the errors are drawn from Gaussian distribution can not be rejected for all the data sets discussed in the present paper.


\normalem
\begin{acknowledgements}
Meghendra Singh thanks DMRC for Support, Shashikant Gupta thanks Tarun Deep Saini for discourse; and compeers of Amity School of Applied Science for eternal assistance. Authors thank to the anonymous reviewer for valuable suggestions.
\end{acknowledgements}

\bibliographystyle{raa}
\bibliography{bibtex}

\end{document}